\documentclass{elsart}
\usepackage{epsfig}
\usepackage{dcolumn}
\usepackage{amssymb}
\def\la{\hbox{{\lower -2.5pt\hbox{$<$}}\hskip -8pt\raise-2.5pt\hbox{$\sim$}}}
\def\ga{\hbox{{\lower -2.5pt\hbox{$>$}}\hskip -8pt\raise-2.5pt\hbox{$\sim$}}}

\begin{document}
\runauthor{Tasitsiomi, Gaskins, and Olinto}
\bibliographystyle{elsart-num}

\begin{frontmatter}

\title{Gamma-ray and synchrotron emission from neutralino annihilation  in the Large Magellanic Cloud}

\author[astro,cfcp]{Argyro Tasitsiomi\thanksref{email}},
\author[physics,cfcp]{Jennifer Gaskins\thanksref{email2}}, 
\author[astro,cfcp]{Angela V. Olinto\thanksref{email3}}
\address[astro]{Department of Astronomy and Astrophysics, and} 
\address[physics]{Department of Physics, and}
\address[cfcp] {Center for Cosmological Physics, 
The University of Chicago, Chicago IL 60637}
\thanks[email]{iro@oddjob.uchicago.edu}
\thanks[email2]{gaskins@oddjob.uchicago.edu}
\thanks[email3]{olinto@oddjob.uchicago.edu}

\begin{abstract}
We calculate the expected flux of $\gamma$-ray and radio emission from the LMC due to neutralino annihilation. 
Using rotation curve data to probe the density profile and assuming a minimum disk, 
we describe the dark matter halo of the LMC using models predicted by N-body simulations.  
We consider a range of density profiles including the NFW profile, a modified NFW profile proposed by Hayashi et al.\ (2003) 
to account for the effects of tidal stripping, and an isothermal sphere with 
a core. We find that the $\gamma$-ray 
flux expected from these models may be detectable by GLAST for a significant part 
of the neutralino parameter space.  The prospects 
for existing and upcoming Atmospheric Cherenkov Telescopes (ACTs) are less optimistic, 
as unrealistically long exposures are required for detection. 
However, the effects of adiabatic compression due to the baryonic component may improve the chances for detection 
by ACTs. The maximum flux we predict is well below EGRET's measurements and thus  EGRET does not constrain the parameter space.  The expected  synchrotron emission generally lies 
below the observed radio emission from the LMC in the frequency range of 19.7 to 8550 MHz.  As long as  $\langle \sigma v \rangle < 2 \times 10^{-26}$cm$^3$s$^{-1}$ for a neutralino mass of 50 GeV, 
the observed radio emission is not 
primarily due to neutralinos and is consistent with the assumption that the main source is cosmic rays.  
We find that the predicted fluxes, obtained by integrating over the entire LMC, are not very strongly dependent on the inner slope of the halo profile, 
varying by less than an order of magnitude for the range of profiles we considered.
 
\end{abstract}


\end{frontmatter}


\section{Introduction}
Decades of observational evidence indicate that galaxies
are surrounded by massive dark matter halos.  It is also
well-established that the baryon fraction of the universe
is too small to account for the measured dark matter density.
Most recent studies suggest that $\sim 23\%$  of the energy density of the universe is in the form of non-baryonic cold dark matter  (CDM) (see, e.g., ~\cite{spergel_etal03}). 

Of the many candidate dark matter particles that have been proposed, the most popular is the Lightest Supersymmetric
Particle (LSP).  Strongly motivated by theoretical considerations,
the LSP is stable in R-parity conserving models and has the appropriate relic density for masses  $m_{LSP} \sim 100$ GeV (see, e.g., \cite{jungman_etal96}).  In  Minimal Supersymmetric extensions of the Standard Model (MSSM), the LSP 
is most likely the lightest neutralino, $\chi$~\cite{ellis_etal84}. Under certain assumptions, 
WMAP results favor
 $m_{\chi} < 500$ GeV \cite{ellis_etal03}, while 
accelerator experiments constrain $m_{\chi}$ $\ga$ $50$ GeV~\cite{hagiwara_02}.

Neutralinos may be detected directly in low background experiments or indirectly by observations of their annihilation
products.  Products of $\chi\bar{\chi}$ annihilation rapidly
decay into neutrinos and $\gamma$-rays, as well as electrons and positrons which emit synchrotron radiation in local magnetic fields.
The rate of $\chi\bar{\chi}$ annihilation depends on the square
of the neutralino number density.  As such, the largest fluxes are
expected to come from the densest regions.  The galactic center is thus an
obvious candidate for its density and proximity~\cite{berezinsky_etal92,gondolo_silk99,cesarini_etal03}, but the numerous 
associated background signals might make it 
difficult to disentangle the neutralino annihilation flux.

High resolution N-body simulations of CDM halos find an abundance of substructure in galaxy-size halos. These studies predict that satellites of our galaxy, such as dwarf spheroidal galaxies, are embedded in dark matter halos of their own. In addition, these simulations find numerous dark ``clumps''.  The detection of neutralino annihilation products from these substructures may  provide information about the clump density profiles as well as neutralino parameters.
Previous studies have considered the annihilation signal from 
dwarf spheroidal galaxies~\cite{baltz_etal00,tyler_02,vassiliev_03} 
and dark matter clumps~\cite{bergstrom_etal99,calcaneo-roldan_moore00,tasitsiomi_olinto02,ullio_etal02,berezinsky_etal03,blasi_etal03,taylor_silk03}. The detectability due to neutralino annihilation of other systems such as the LMC~\cite{gondolo_94}, M31~\cite{falvard_etal02}, and M87~\cite{baltz_etal00}, the globular cluster Palomar 13~\cite{giraud_etal02}, and galaxy 
clusters~\cite{colafrancesco_mele01} have also been studied.

At a distance of 50.1 kpc~\cite{vandermarel_etal02}, the LMC is an
obvious choice for neutralino detection. There are several 
LMC mass estimates in the literature (e.g., 
\cite{meatheringham_etal88,schommer_etal92,kim_etal98,sofue_99,alves_nelson00,vandermarel_etal02}), 
and most agree within error bars.  According to the most recent study, the mass of the LMC enclosed in a radius of $8.9$ kpc is $M_{LMC} \simeq (8.7 \pm 4.3) \times 10^9 
M_{\odot}$. To study neutralino annihilation it is important that the LMC be dark matter dominated.  
Van der Marel et al.~\cite{vandermarel_etal02} found that more than half of the LMC  mass is in a dark halo. 
In addition, Sofue~\cite{sofue_99}  argued that the LMC must have a dark and compact bulge with an  anomalously high mass-to-light ratio $M/L \sim 20-50$ $M_{\odot}/L_{\odot}$.

The LMC has been observed in a wide range of frequencies, providing a wealth of data to draw upon. Past estimates of the LMC neutralino signal~\cite{gondolo_94} focused solely on $\gamma$-rays and did not fit modern density profiles to the LMC halo. New rotation curve data and recent theoretical models of tidally stripped dark matter sub-halos motivate a new estimate of the LMC neutralino flux in $\gamma$-rays and in synchrotron emission. 
In addition, next-generation Atmospheric Cherenkov 
Telescopes (ACTs) and the upcoming GLAST satellite make this study of the 
LMC even more timely.  
 
 The flux of both $\gamma$-ray and synchrotron emission from neutralino annihilation in the LMC depends on the square of the density profile of the LMC dark halo. Thus,  we begin the next section with a discussion of
 the structure of the LMC and describe fits of different dark matter profiles to the LMC rotation curve data. In \S III we calculate the $\gamma$-ray flux and investigate the detectability prospects for ACTs and GLAST\@.  In \S IV we calculate the synchrotron flux and compare it to observations. We conclude in \S V.

\section{Modeling the LMC dark matter halo}
Rotation curve data provides a dynamical measure of the mass
distribution via
\begin{equation}
V_{rot}(r)=\sqrt{\frac{GM(r)}{r}}\ , 
\end{equation}
where $V_{rot}(r)$ is the rotation velocity at radius $r$ and 
$M(r)$ is the total mass interior to $r$, $M(r)=\int_0^r \rho(r^{'}) d^3r^{'}$. 
Given an observed rotation curve, we can study how well a particular model of the density distribution fits the observations. For the LMC rotation curve data, we use HI data from  Kim et al.~\cite{kim_etal98} that spans the range from 0.05 to 3.09 kpc with a velocity resolution of $\sim 1.65$ km/s,  and carbon star  data 
from Alves and Nelson~\cite{alves_nelson00}  that covers a radius range of 4 to 8.2 kpc.  We assume $8 \%$ error bars for the HI data given in \cite{kim_etal98}.

Numerous N-body simulations have focused on understanding the 
structure of dark matter halos.  Navarro, Frenk and White (NFW)~\cite{navarro_etal95,navarro_etal96}
performed N-body simulations of CDM halos and found that
for a large range of masses the density profiles
 were well-described by the simple formula
\begin{equation}
\rho(r)=\frac{\rho_{o}}{\left({r / r_{s}}\right) \left({1 + {r
/ r_{s}}}\right)^{2}}\ , 
\end{equation}
with $\rho_{o}$ and  $r_{s}$ a characteristic density and scale radius, 
respectively. 
The NFW profile can be alternatively characterized by 
$r_{s}$ and the concentration
parameter $c$ which is related to $\rho_o$ and is defined 
as the ratio $r_{v}/r_{s}$ where  $r_{v}$ is the virial radius. The virial radius is defined as 
the radius within which a certain virial 
overdensity is reached, typically 180 times the average density of the universe. 
Behaving as $r^{-1}$ at the center, the NFW profile  is shallower than both the singular 
 isothermal sphere and the Moore et al.~\cite{moore_etal98} density profiles
which behave at the center as $r^{-2}$ and $r^{-1.5}$, respectively.
Given the apparent discrepancy between the predictions of N-body simulations 
and observations (see, e.g.,~\cite{tasitsiomi_02}), with the latter favoring shallow rather than very steep profiles at the center, we primarily focus on the NFW profile. This is a conservative assumption for CDM profiles since steeper density profiles lead  to larger fluxes from neutralino annihilation. 
In particular, the effect of the baryonic component on the central density profile 
may increase the central density by more than an order of magnitude (see, e.g., \cite{prada_etal04}).

Hayashi et al.~\cite{hayashi_etal03} performed N-body simulations to investigate how the density profile of an NFW halo changes due to tidal stripping.  
They found that the resulting profile can be described by a simple modification to the NFW profile:
\begin{equation}
\rho(r)=\frac{f_{t}}{1 + (r / r_{te})^{3}} \: \rho_{_{NFW}} .
\end{equation}
The factor $f_{t}$ is a dimensionless measure of the reduction in central
density while the ``effective'' tidal radius $r_{te}$ describes
the outer cutoff due to tides. The effect of tidal stripping on the initial NFW profile is to lower the characteristic density, to make it 
steeper in the outer region, and to introduce an outer cutoff.
Since there is evidence that the LMC may be undergoing tidal disruption, we also model its halo using this profile. 

As observations find that more than half of the mass 
within $\sim 9$ kpc of the LMC is dark~\cite{vandermarel_etal02},  we assume that the disk is minimal  (i.e., that the gravitational potential of the LMC is entirely due to dark matter) and search for the best fit NFW 
and Hayashi et al.\ profiles. For the NFW profile we find $r_s=9.16$ kpc and 
$\rho_{o}=1.66 \times 10^7 M_{\odot}/ \rm{kpc}^3$,  corresponding to a concentration index $c \simeq 11.2$, 
and a $\chi^2$ per degree of freedom of 1.01.  

For the purposes of fitting the Hayashi et al.\ profile, we absorb the factor
$f_{t}$ into the NFW parameter $\rho_{0}$ and call the resulting parameter
$\tilde{\rho_{0}}$.  To distinguish between different models with almost 
identical $\chi^2$ per degree of freedom values, we first fix the scale radius at the NFW best-fit
value, $r_{s}=9.16$ kpc.  In this case, we find $r_{te}=6.12$
kpc, $\tilde{\rho_o}= 8.18 \times 10^6 M_{\odot}$/kpc$^{3}$ with a $\chi^2$ per degree of freedom equal to 0.91.  

Recent N-body simulations (\cite{kravtsov_03}, see also 
\cite{hayashi_etal03,klypin_etal99}) suggest that the process of tidal disruption
of an NFW halo of mass comparable to the LMC may lead to a decrease in the scale radius of up to about 30\%.  The
density profile of the LMC at the present is well-fitted by $r_{s}=9.16$ kpc, so we can assume that this value represents a decrease of not more than 30\% from the original value. Thus we can extrapolate that the scale radius of the original
NFW halo that gave rise to the LMC was not more than $\sim$ 13.1 kpc ($\simeq 9.16/0.7$ kpc $\simeq 1.4 \times 9.16$ kpc).  
To understand the effect that a larger scale radius has on the Hayashi et al. density profile and the resulting flux,  we again fit the rotation curve data using a fixed $r_{s}=13.1$ kpc.  The resulting 
best-fit Hayashi et al.\ profile parameters are given by
$r_{te}=4.97$ kpc, $\tilde{\rho_o}= 5.46 \times 10^6  M_{\odot}$/kpc$^{3}$ 
with a $\chi^2$ per degree of freedom equal to 0.81.  

\begin{figure}[htb]
\centerline{\psfig{figure=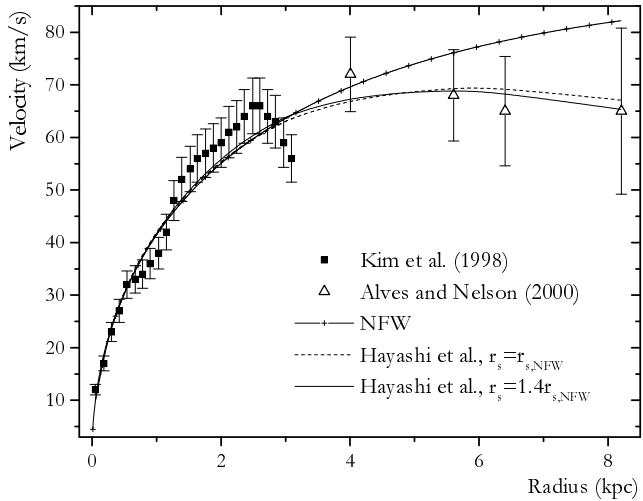,width=4.2in}}
\setlength{\unitlength}{0.0105in}%
\caption{Rotation curve of the LMC: square points from
HI data~\cite{kim_etal98} and triangle points from carbon star data~\cite{alves_nelson00}. The crossed curve is the NFW fit. 
The dashed curve is the Hayashi et al.\ fit obtained by fixing the scale radius to the
best fit value found for the NFW profile.  The solid line corresponds to the Hayashi et al.\ fit obtained by fixing the scale radius at $r_{s} \sim 1.4 r_{s,NFW}$ to 
account for the possible evolution of $r_{s}$ (see text for details).}
\label{fig:rc}
\end{figure}

All three fits to the rotation curve data are shown in Fig.\ \ref{fig:rc}.
While all the fits are acceptable, the Hayashi et al.\ profiles better fit the data
at the outer radii.  The masses predicted by these models are in reasonable
agreement with estimates in the literature -- the Hayashi et al.\ gives
a mass of  $\sim 9 \times 10^9 M_{\odot}$ within $8.9$ kpc, while the NFW fit
gives $\sim 3 \times 10^{10} M_{\odot}$ within 8.9 kpc, slightly above the observationally determined $(8.7 \pm 4.3) \times 10^9  M_{\odot}$ within 8.9 kpc by~\cite{vandermarel_etal02}.  

The relatively small tidal radii of the Hayashi et al.\ fits deserve comment.
Observations typically indicate the tidal radius for the LMC $r_t$ is greater than about 10 kpc (\cite{vandermarel_etal02} and references therein). 
However, different observational methods for determining this quantity may
give significantly different values.
While a precise determination of the tidal radius would require knowing the distance at which regular isopleths of a certain class
of stars end, more practical estimators involve the Roche limit (valid for point masses), or the Jacobi limit (where the centrifugal force on the satellite is taken into account).
For instance, the tidal radius determined via the Roche limit
($r_{tR}=(M_{LMC}/2 M_{MW})^{1/3} d$)
relies upon approximate mass estimates for both the Milky Way halo ($M_{MW}$) and the LMC. These methods do not directly 
correspond to the  Hayashi et al.\ profile 
parameter $r_{te}$, which is an  \emph{effective} tidal radius.  Careful inspection  
of the enclosed mass as a function of radius for tidally 
stripped objects in  the Hayashi et al.\ study reveals 
that the mass does not necessarily reach a plateau at $r=r_{te}$, but continues 
to increase with radius (e.g., up to 2 times $r_{te}$, which is closer 
to the observational  $r_t $).  Consequently the parameter $r_{te}$ does not necessarily 
correspond to the observationally determined tidal radius.  Lastly, even 
if we force $r_{te}$= 15 kpc, we find that the Hayashi et al.\ fit becomes 
essentially the same as the NFW best fit (the Hayashi et al.\ best fit $r_s=9.35$ kpc 
while the NFW best fit $r_s=9.16$ kpc). This is not surprising given 
that with such a large $r_{te}$ the difference between the 
Hayashi et al.\ and the NFW profiles arises at radii much 
larger than that for which there is  rotation curve data available.
 
For completeness, we consider three additional profiles.  We examine two 
recently proposed profiles:  the Stoehr et al.~\cite{stoehr_etal02} profile 
which was found to accurately fit rotation velocity curves of 
satellite halos in some simulations, and the more centrally 
concentrated Moore et al.~\cite{moore_etal98} profile.  We also look 
at a shallower profile -- an isothermal sphere with a core as derived in \cite{alves_nelson00}
by fitting the same rotation curve data we use here.
Both  the Moore et al.~\cite{moore_etal98} and the 
Stoehr et al.~\cite{stoehr_etal02} profiles 
seem less favored by the rotation curve data -- they give $\chi^{2}$ values per degree of freedom 
5.49 and 4.24 respectively.
The  isothermal sphere with a core represents a less ideal scenario.  
Instead of making the assumption of a minimum disk, the stellar 
disk and gas contributions are included in the model  by Alves and Nelson~\cite{alves_nelson00}. In this case, the halo profile is given by 
\begin{equation}
\rho(r)= \frac{\rho_{o}}{1+ (\frac{r}{a})^2}\ ,
\label{core}
\end{equation}
with $\rho_{o} = 10^8 M_{\odot}/$kpc$^3$ and  core radius $a=1$ kpc.

\section{The $\gamma$-ray Emission}
Neutralino annihilation proceeds through a number of channels, many of which
produce $\gamma$-rays in the final state (see e.g., \cite{cesarini_etal03}).  
Here we consider only the continuum $\gamma$-ray emission due to the decay of neutral pions produced in  hadronic jets.  
\begin{eqnarray}
\label{anni}
\chi+\bar{\chi} & \rightarrow & \pi^{0}+X   \nonumber  \\
                &             & \hookrightarrow \pi^{0} \rightarrow 
\gamma+\gamma.
\end{eqnarray}
Given the uncertainties in the dark matter density distribution, a full calculation of the gamma-ray emission in the large dimensional parameter space of SUSY models is not warranted. Instead, we use the approximate Hill spectrum~\cite{hill_83} based on the leading-log approximation (LLA).  For MSSM models,  this approximation is reasonable for neutralino masses below the W mass and above the top quark mass. Furthermore, the simplified spectra derived below vary from the fit obtained in~\cite{bergstrom_etal01} by no more than 
a factor of a few for neutralino masses from 10 GeV up to a few TeV. 

Assuming that the hadronic jets contain only pions and that equal numbers of $\pi^{+}$'s, $\pi^-$'s, and $\pi^{0}$'s are produced,
the neutral pion spectrum is given approximately by
\begin{equation}
\label{spec1}
\frac{dN_{\pi^{0}}}{dx_{\pi}} \simeq \frac{5}{16} x_{\pi}^{-3/2}
(1-x_{\pi})^{2}
\end{equation}
with $x_{\pi}=E_{\pi^{0}} / m_{\chi}$.
Furthermore, the probability per unit energy that a neutral pion with 
energy $E_{\pi^{0}}$  produces a photon  with energy $E_{\gamma}$
through the process shown in Eq.\ (\ref{anni}) is $2/E_{\pi^{0}}$.
Thus, from Eq.\ (\ref{spec1}) we get for the continuum
photon spectrum,
\begin{equation}
\frac{dN_{cont.}}{dx_{\gamma}} \simeq \int_{x_{\gamma}}^{1} \frac{2}{y}
\frac{dN_{\pi^{0}}}{dy} dy
\end{equation}
with $x_{\gamma}={E_{\gamma}/m_{\chi}}$ and $y={E_{\pi^{0}}/m_{\chi}}$. 
The number of continuum photons produced per annihilation 
through this channel with
energy greater than a specified threshold is found by integrating this
spectrum, which gives
\begin{eqnarray}
\label{ngammaeqn}
N_{cont.}(E_{\gamma} \ge E_{th}) =
  \int_{x_{th}}^{1} \frac{dN_{cont.}}{dx_{\gamma}}dx_{\gamma} \\
=
\frac{5}{6}x_{th}^{3/2}-\frac{10}{3}x_{th} + 5\sqrt{x_{th}}
+\frac{5}{6 \sqrt{x_{th}}} -\frac{10}{3} 
\end{eqnarray}
where  $x_{th}={E_{th}/m_{\chi}}$.

The $\gamma$-ray emission coefficient $j$ is
\begin{equation}
\label{gammas_emission}
j=N_{\gamma}(E \geq E_{th}) \frac{\langle \sigma v \rangle_{\gamma}}{m_{\chi}^2} \rho^2(r)\ ,
\end{equation}
where $N_{\gamma}$ is the number of photons above the energy  threshold
as found in Eq.\ (\ref{ngammaeqn}).  The quantity 
$\langle\sigma v\rangle_{\gamma}$ is the thermally averaged cross section times
velocity for annihilation
into $\gamma$-rays, and
$\rho(r)$ is the density profile of the halo. 
The specific intensity of $\gamma$-rays, $I(R)$, as a function of projected
radius $R$, is given by
\begin{equation}
I(R)=\frac{1}{4 \pi} \ 2 \int_{0}^{z_{max}(R)} j dz \ ,
\end{equation}
for which  $z$ is  the coordinate along the line of sight and 
$z_{max}(R)=\sqrt{r_{max}^{2}-R^{2}}$ where $r_{max}$ 
is the radius of the object. 
The $\gamma$-ray flux $F$  from an  object  
at a distance $d$ from us is given by 
\begin{equation}
\label{fluxeqn}
F=\frac{1}{2}\frac{1}{4{\pi}d^2} \int_{0}^{r_{max}} j  d^{3}r \ . 
\end{equation}

The radial dependence of $j$ is confined to $\rho^2(r)$.  We  isolate the 
dependence of the flux on the specific halo profile
and distance to the LMC by defining $K$ as 
\begin{equation}
K=\frac{1}{2}\frac{1}{4{\pi}d^2} \int_{0}^{r_{max}} \rho^{2}(r) d^{3}r\ .
\label{K}
\end{equation}
The $\gamma$-ray flux is then given by
\begin{equation}
F=K \frac{N_{\gamma} \langle \sigma v \rangle_{\gamma}}{m_{\chi}^2}\ .
\label{fluxk}
\end{equation}
For $r_{max}$  we use 3.1 kpc,  which is the radial extent
of the LMC as observed in $\gamma$-rays by EGRET~\cite{sreekumar_etal92}.
We take the distance to the LMC to be 50.1 kpc as found 
in~\cite{vandermarel_etal02}.  
We give $K$ in units of (GeV/c$^{2}$)$^{2}$/{cm}$^5$, which
gives the flux $F$ in cm$^{-2}$ s$^{-1}$ with $\langle\sigma v\rangle_{\gamma}$ in
cm$^{3}$ s$^{-1}$ and $m_{\chi}$ in GeV/c$^2$.

For the NFW model $K = 3.65 \times 10^{19}$ (GeV/c$^{2}$)$^{2}$/{cm}$^5$,
 while for the Hayashi et al.\ fits
$K = (0.84 - 0.85) \times 10^{19}$ (GeV/c$^{2}$)$^{2}$/{cm}$^5$
 depending on the exact value of the scale
radius.  
For the isothermal sphere with a core we obtain $K=0.43 \times 10^{19}$ (GeV/c$^{2}$)$^{2}$/{cm}$^5$.
The more centrally concentrated Moore et al.\ and the Stoehr et al.\ profiles give slightly higher fluxes, 
by factors $\sim 2-4$ with respect to the NFW case.
Clearly, the small variations between the fluxes produced by 
these models are overwhelmed by observational uncertainties.  
Fits to all of these profiles give
$K$ on the order of $10^{19}$ (GeV/c$^{2}$)$^{2}$/{cm}$^5$.
These profiles behave similarly and
consequently give similar fluxes, thus, for simplicity we focus on the NFW profile.
The mild dependence on the halo profile, and more specifically on its central slope, is expected given that we calculate
the flux from the entire observed extent of the LMC, rather than just from the innermost region. 
Thus, the 
flux  depends most strongly  on the ``normalization'' of the profile 
which is set by the mass 
enclosed in the volume over which we integrate.

EGRET detected a flux of  $(14.4 \pm 4.7) \times 10^{-8}$  
photons $(E > 100\ \rm{MeV})$ cm$^{-2}$ s$^{-1}$ from  the LMC~\cite{hartman_etal99}.  
Using our fit to the NFW profile and an energy threshold of 100 MeV,
the maximum flux produced by a viable SUSY model is $\simeq 3.3 \times 
10^{-9}$ photons cm$^{-2}$ s$^{-1}$, corresponding to $m_{\chi} 
\simeq 50$ GeV and $\langle \sigma v \rangle_{\gamma}
\simeq 2 \times 10^{-26}$ cm$^3$ s$^{-1}$.  
While being consistent with the 
flux detected by EGRET, even the maximum predicted flux is
almost  two orders of magnitude too low, suggesting that the 
primary source is cosmic rays.  Consequently, EGRET's observations do not constrain the parameter space.

EGRET's measurement indicates that cosmic ray induced $\gamma$-rays may be an additional background
component to consider when trying to detect flux from neutralino annihilation. Following a method similar to that presented 
in ~\cite{pavlidou_fields01}, we calculate this cosmic ray induced background.
We review very briefly here the basic points of the calculation and refer the reader to that study for more
details.  The  basic assumption 
is that cosmic ray acceleration takes place in supernovae remnants.  Working in the frame of
the ``leaky box model'' for cosmic ray propagation, the authors calculate
the expected cosmic ray flux in the LMC by considering the cosmic ray flux measured 
for the Milky Way and the supernova rates observed for the Milky Way and the LMC.\    
For the $\gamma$-ray emissivity they include $\gamma$-rays from decays of neutral pions produced
by proton and heavier nuclei interactions, as well as $\gamma$-rays produced via bremsstrahlung
of cosmic ray electrons. 

\begin{figure}[t]
\vspace{2cm}
\centerline{\epsfxsize=9cm \epsfysize=9cm \epsffile{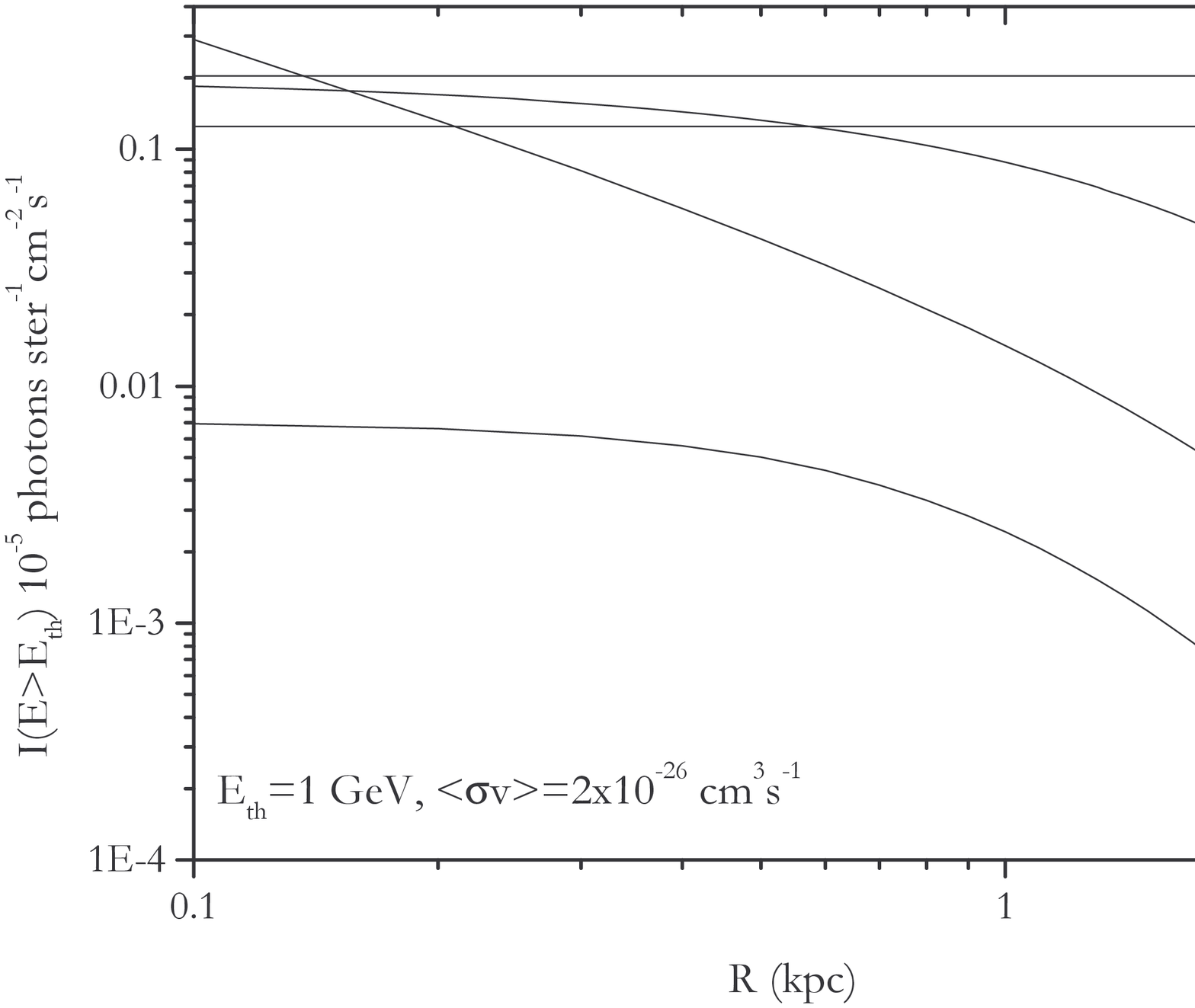} \hspace{-1cm} 
\epsfxsize=9cm \epsfysize=9cm \epsffile{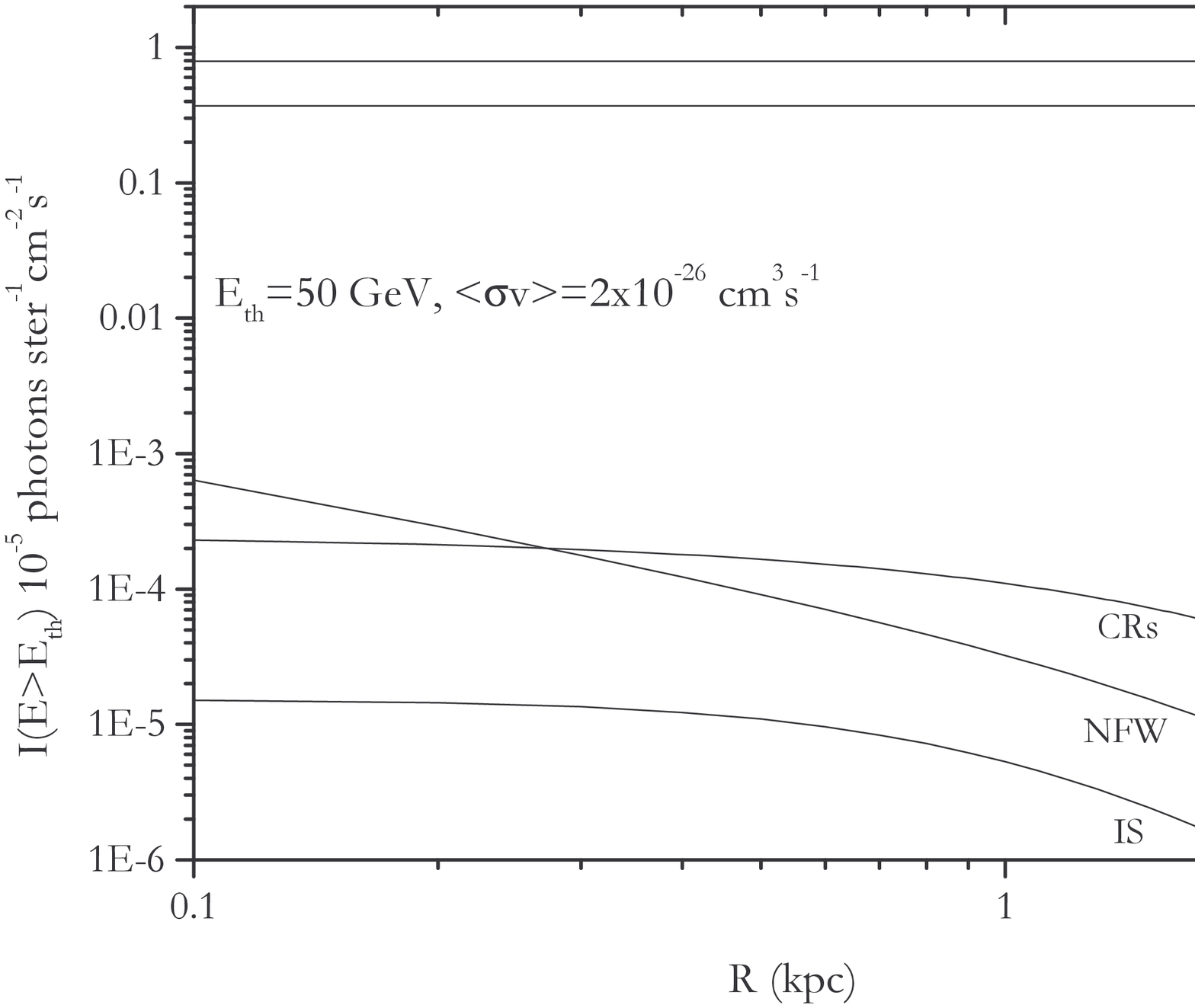}}
\vspace{-3cm}
\caption{{\it Left panel:} Specific intensity as a function of projected distance from the 
center of the LMC for several background contributions and the neutralino signal above 1 GeV.\
The signal is shown for the NFW profile and the isothermal sphere with a core (IS).
The background components are labeled as follows: G for galactic, EG for extragalactic, CRs for cosmic rays.
{\it Right panel:} Same as left panel but for an energy threshold of 50 GeV. The background components
in this case are labeled as follows: H for hadronic and E for electronic. }
\label{fig:intensity_comp}
\end{figure}

The calculation of the cosmic ray induced $\gamma$-rays in \cite{pavlidou_fields01} focuses
on the flux coming from the entire LMC.\ 
However, from an observational standpoint it is useful to examine the angular dependence of the signal.  
We calculate the specific intensity as a function of projected distance from the center of the LMC.\ To do this, it is necessary to
assume a certain distribution for the HI and H$_{2}$ gas in the LMC, 
rather than using a mean gas surface density as
in \cite{pavlidou_fields01}. 
According to  \cite{kim_etal98} , the gas in the LMC appears to be distributed approximately 
as a disk with a $z$-scalelength of about $600$ pc and the HI disk is about $7.3$ kpc in diameter.  
Assuming an exponential distribution in $R$ as well, we equate the $7.3$ kpc  to 6 radial
scalelengths and obtain  a scalelength of $\sim 1.2$ kpc, which is close to 
the scale length of the stellar distribution (see, e.g., \cite{alves_nelson00}).  
Lastly, we normalize the density profile so that when integrated  
over the line of sight coordinate, $z$,  and  the projected radius, $R$, 
it yields a gas mass in agreement with observations. 
(Note that because the LMC is viewed approximately face-on, 
the rotation axis corresponds approximately to the line of sight direction and here we use $z$ to describe both.)

In Figure \ref{fig:intensity_comp} we show a comparison of the specific 
intensity of the different backgrounds, as well as that 
for the NFW profile and the isothermal sphere with a core.
The left panel assumes an energy threshold of 1 GeV which is appropriate for GLAST, 
while the right panel assumes a 50 GeV threshold appropriate for ACTs.
For GLAST  the relevant backgrounds are 
the galactic and extragalactic
diffuse emission.  We use the expression found in~\cite{sreekumar_etal98} 
for the isotropic 
extragalactic background and the expression from~\cite{bergstrom_etal98} 
evaluated at the LMC
galactic coordinates ($l=280.46^{o}, b=-32.89^{o}$) 
for the galactic background. For ACTs the relevant  backgrounds 
are the hadronic and electronic cosmic ray shower 
contributions~\cite{bergstrom_etal98,longair_92}, though it is worth noting that ACTs can reject
hadronic showers with high efficiency.
Clearly, 
the cosmic ray background is not dominant at these energy thresholds. 
At very low energy thresholds ($100$ MeV -- not shown in the Figure) the cosmic ray background 
does become dominant in the very
central regions (inner $\sim 0.5$ kpc). As a consistency check, after calculating the specific intensity
of the cosmic ray induced flux 
above 100 MeV, we integrate it over solid angle and recover the $\sim 10^{-7}$ photons cm$^{-2}$ s$^{-1}$  prediction of 
\cite{pavlidou_fields01}.  This value is approximately equal to the flux measured by EGRET.\ 

\begin{figure}
\vspace{2cm}
\centerline{\epsfxsize=9cm \epsfysize=9cm \epsffile{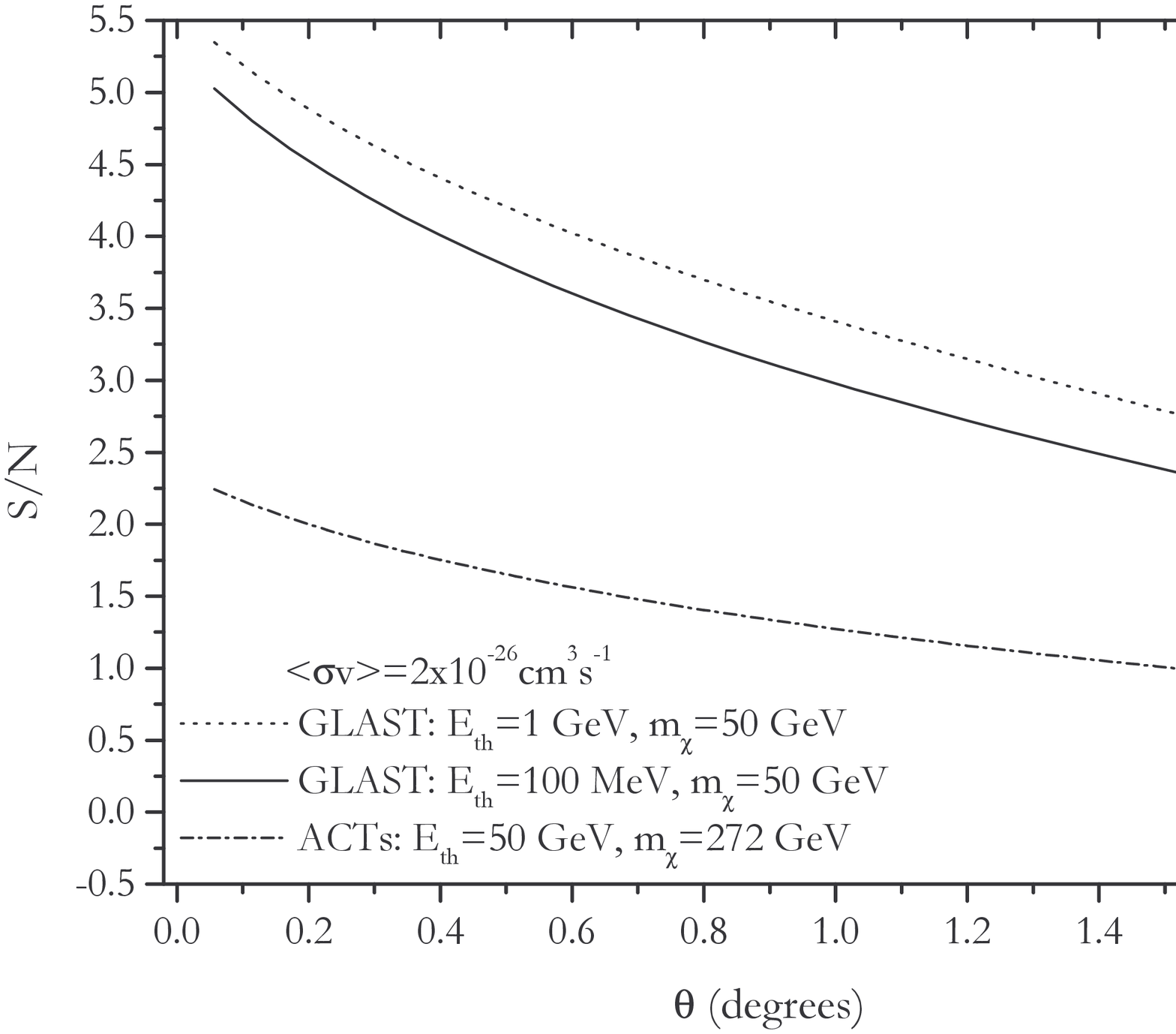}}
\vspace{-3cm}
\caption{Angular dependence of the signal-to-noise ratio for a NFW dark matter halo profile.
For GLAST we use $A_{eff}=10^{4}$cm$^{2}$ and 1 year of observation.
For ACTs we take $A_{eff}=10^{9}$cm$^{2}$ and 1 month of observation.}
\label{fig:s2n}
\end{figure}

Observationally, the most relevant quantity is the signal-to-noise ratio and its  angular and
energy dependence. In Figure~\ref{fig:s2n} we present 
the signal-to-noise for the flux within an angle $\theta$ , assuming the NFW profile.  We show results for 
three different energy thresholds with the neutralino parameters for each chosen to optimize the signal.
We use the specifications for each instrument as indicated.
For the noise all the relevant backgrounds, including the cosmic ray
induced emission, were used. From the figure it is clear that focusing on the central regions may help 
to achieve higher
signal-to-noise ratios. As expected, for the isothermal sphere with a core (not shown here),
the results are less optimistic.

\begin{figure}[htb]
\vspace{2cm}
\centerline{\psfig{figure=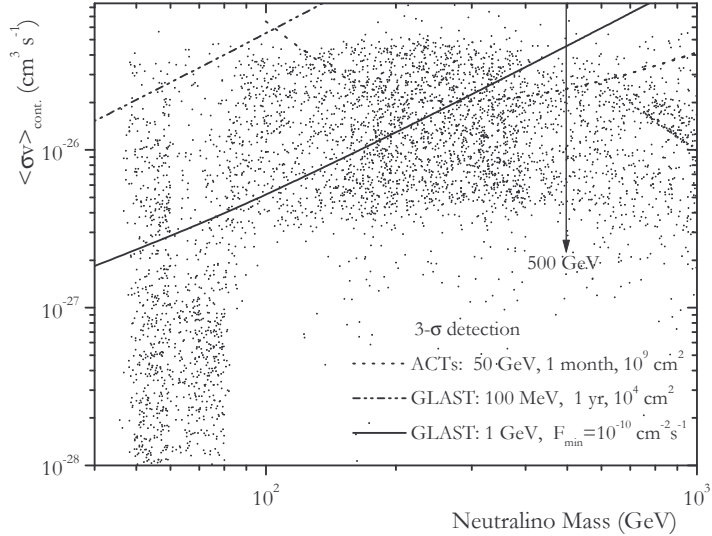,width=4.5in}}
\setlength{\unitlength}{0.0105in}%
\vspace{-3cm}
\caption{The minimum detectable $\langle \sigma v \rangle_{\gamma}$ versus $m_{\chi}$ for the NFW profile. SUSY models above each
curve yield a detectable signal for the instrument and observational 
parameters assumed. The dotted line represents observations by upcoming 
ACTs with 50 GeV threshold, 100 hours of observation, and $10^8$ cm$^2$ effective area. The 
dashed line shows detectability by ACTs with an effective area of $10^9$ cm$^2$ 
which will be achievable only at high energy thresholds $(\sim 1$ TeV). The solid line is 
derived for GLAST using the expected flux sensitivity $\sim 10^{-10}$ cm$^{-2}$ s$^{-1}$.  The dot-dashed 
line is derived for GLAST for 1 year of observation, an effective area of $\sim 10^4$ cm$^2$, and
an energy threshold of 100 MeV. }
\label{fig:susy}
\end{figure}

Taking the idealized case of no systematic errors, the detectability condition
requires that the
signal-to-noise exceed the desired significance $s$,
\begin{equation}
\frac{F_{min} A_{eff} t}{\sqrt{N_{b}}} \geq s\ ,
\label{fmin}
\end{equation}
where $F_{min}$ is the minimum required flux for an $s$-$\sigma$ detection 
to be achieved
by an instrument of effective area $A_{eff}$, integration time $t$, 
and number of background counts
for the exposure $N_{b}$.  
The detectability of the neutralino induced flux for a 
particular SUSY model depends on the neutralino mass
and the annihilation cross section for that model.
The neutralino induced flux must exceed $F_{min}$ in order to be 
detectable.  This condition can be used to divide the $m_{\chi}$-$\langle
 \sigma v \rangle_{\gamma}$ parameter space into detectable and undetectable models
for each instrument. 
This is shown in Fig.\ \ref{fig:susy}. The points represent SUSY models produced using the DarkSUSY package~\cite{DarkSUSY}. Models with $m_{\chi}$-$\langle
 \sigma v \rangle_{\gamma}$ above a given line are accessible to observation by the corresponding
instrument, while models in the lower region do not yield a detectable flux. 

For GLAST, the solid line is derived for the optimistic case 
of a flux sensitivity of $\sim 10^{-10}$ photons cm$^{-2}$ s$^{-1}$,  which is the expected point source flux sensitivity of GLAST for energies above
$\sim$ 1 GeV~\cite{deangelis_00}. This flux sensitivity  corresponds to a 5-$\sigma$ detection and 1 year of on-target observation.  During normal operation, each source will only be in GLAST's field of view about 20\% of the time.  Consequently, the actual time required to achieve a detection of the same significance is approximately 5 times the necessary on-target time.  This is within reach since the GLAST science requirement is 5 years of operation, 
while the lifetime goal is 10 years.
If GLAST achieves its expected sensitivity, 
then it  will be able to detect the neutralino signal for a significant 
portion of the parameter space. This is particularly true in light of the recently 
derived limit $m_{\chi}  < 500$ GeV~\cite{ellis_etal03} 
(vertical arrow in Fig.\ \ref{fig:susy}). Additionally, a typical value for the total thermally averaged cross section 
times velocity is 
$\langle \sigma v \rangle_{tot} \simeq 2 \times 10^{-26}$ cm$^3$ s$^{-1}$ (using 
$\Omega_m h^2 \simeq 3 \times 10^{-27} 
\rm{cm}^3 \rm{s}^{-1}/ { \langle \sigma v \rangle_{tot}}$ from 
e.g.,~\cite{jungman_etal96}, with
$\Omega_m h^2 \simeq 0.135$~\cite{spergel_etal03}).
By making the approximation that
all neutralinos annihilate into pions,
and that all neutral pions (roughly one third of all pions) annihilate to produce $\gamma$s,
$\langle \sigma v \rangle_{\gamma}$ is about a factor of 3 smaller than $\langle \sigma v \rangle_{tot}$.
Thus, a typical value for $\langle \sigma v \rangle_{\gamma}$ is, e.g., 
$\sim 7 \times 10^{-27} \rm{cm}^3 \rm{s}^{-1}$. 
If no annihilation signal is detected, the corresponding part of the parameter
space can be ruled out.  

For an energy threshold equal to 100 MeV we take $A_{eff}=10^4$ cm$^2$, 1 year of on-target
observation and $\Delta \Omega_{LMC} \simeq 1.13 \times 10^{-2}$ sr for the solid angle 
relevant for the collection of noise (since the LMC will be resolved by GLAST).
With the
sensitivity for this energy threshold  only a small 
portion of the parameter space will be detectable (dot-dashed line in Fig.\ \ref{fig:susy}).  
Clearly, 
the best chance for detection is by using moderate energy thresholds, such as the 1 GeV case, 
for which the backgrounds are relatively low but the source photons are still numerous.
To identify neutralino annihilation as the origin of the observed flux,
the spectrum and its characteristic features, such as
the cutoff at $E=m_{\chi}$, may be useful. 
In addition, monochromatic lines produced by neutralino annihilation 
(e.g., the $\gamma \gamma$ line at $E=m_{\chi}$) can be excellent observational signatures for a 
small region of parameter space where these processes are not strongly suppressed (see, e.g.,~\cite{tasitsiomi_olinto02}).

The detectability prospects for existing and upcoming ACTs are less
optimistic.  The commonly assumed 
specifications are  $A_{eff}=10^{8}$ cm$^2$, $E_{th}=50$ GeV and 100 
hours of observation. For these specifications we  find that no models will be detectable.  
A large integration time ($\sim 1$ month) and effective area 
($\sim10^9$ cm$^2$)  (dotted line in the Figure \ref{fig:susy}) would improve the chances for detection. 
However, such integration times are 
fairly long for ACT observations and such large effective areas for an energy threshold  
of $\sim$ 50 GeV are beyond the goals of existing and upcoming ACTs. Effective areas of order 
$10^9$ cm$^2$ are expected to be achieved by ACTs for energies $\sim 1$ TeV, but the number 
of $\gamma$-rays  produced by dark matter annihilation with energies 
$\ga$ 1 TeV is expected to be zero for models where the upper limit on $m_{\chi} < 500$ GeV holds.
 
Of the upcoming ACTs, the LMC will only be visible to
HESS~\cite{hofmann_etal00} and CANGAROO III~\cite{mori_etal00} due to its
position in the sky. However, the zenith-angle dependence of the effective area 
of these instruments greatly decreases the prospects for detection. The minimum zenith angle of 
the LMC (at local meridian crossing) is $\sim$  $46^0$ and $39^0$ for 
HESS and CANGAROO III, respectively. The energy threshold increases and the effective area is drastically reduced for these large zenith angles, requiring even longer integration times.

In this section we focused our flux calculations on the NFW fit and also discussed the less favorable case of the
isothermal sphere with a core.
In addition, we estimated the change in flux for other profiles.  Of the range of profiles we considered, 
the fluxes produced vary by less than an order of magnitude, and hence the detectable cross sections 
and neutralino masses do not differ greatly from those of the NFW profile. This 
mild dependence on the detailed profile arises from our integrating to a radius $R$ which includes most of the LMC 
as it appears in $\gamma$-ray observations. We find that the cosmic ray induced background is largely subdominant 
compared with other background sources, and that for the signal to exceed all backgrounds, observations of the central region of the LMC 
may be the most promising.

\section{The synchrotron emission}

Neutralino annihilation not only generates neutral pions, but also a comparable number of charged pions.
The charged pions decay as
\begin{equation}
\pi^+ \rightarrow \mu^+ \nu_{\mu} \,\, {\rm and} \,\,
\pi^- \rightarrow \mu^- \bar{\nu}_{\mu} \,.
\label{pidecay}
\end{equation}
Muons then decay via 
\begin{equation}
\mu^+ \rightarrow e^+ \bar{\nu}_{\mu} \nu_e \,\, {\rm and} \,\,
\mu^- \rightarrow e^- \nu_{\mu} \bar{\nu}_e \,.
\label{mudecay}
\end{equation}
In the presence of magnetic fields,  the electrons and positrons produced in this way generate synchrotron radiation.  The synchrotron emission from neutralino annihilation has been considered
for a variety of  objects including the galactic center~\cite{berezinsky_etal92}, dark matter clumps~\cite{blasi_etal03}, Draco~\cite{tyler_02}, and 
halos of galaxy clusters~\cite{colafrancesco_mele01}.
Here we follow previous studies and calculate the synchrotron emission due to neutralino annihilation in the LMC.

The radio flux from the LMC has been measured in frequencies
from 19.7 to 8550 MHz. The electron energies with maximum synchrotron emission in this range  are relatively low ($\sim 20$ GeV for the most energetic 
8550 MHz synchrotron photons for a 5 $\mu$G magnetic field).
Thus, we can consider only the dominant source of electrons and positrons for these energies:
the $\pi^{\pm}$ decays.  This is valid for gaugino-like neutralinos (see \cite{colafrancesco_mele01} and references
therein).  We use the analog of Eq. (\ref{spec1}) (substituting $\pi^{0}$ with charged pions which we denote by
$\pi$) for the charged pion spectrum, $dN_{\pi}/dE_{\pi}$,  
produced by quark fragmentation.

The number of electrons and positrons produced per annihilation per energy interval $dN_e/dE_e$ is then given by 
\begin{equation}
{dN_e \over dE_e} = \int_{E_e}^{m_{\chi} c^2}
\int_{E_{\mu}}^{aE_{\mu}} \frac{dN_{\pi}}{dE_{\pi}}
{dN_{\mu}^{(\pi)} \over dE_{\mu}}
{dN_e^{(\mu)} \over dE_e}\,dE_{\pi}\,dE_{\mu}\,,
\label{fragtot}
\end{equation}
where $a \equiv (m_{\pi} / m_{\mu})^2$,
\begin{equation}
{dN_{\mu}^{(\pi)} \over dE_{\mu}} =
{1 \over E_{\pi}}
{m_{\pi}^2 \over m_{\pi}^2 - m_{\mu}^2} \ ,
\label{fragpi}
\end{equation}
and
\begin{equation}
{dN_e^{(\mu)} \over dE_e} = {2 \over E_{\mu}}
\left[ {5 \over 6} - {3 \over 2} \left( {E_e \over E_{\mu}}
\right)^2 + {2 \over 3} \left( {E_e \over E_{\mu}} \right)^3
\right]\,.
\label{fragmu}
\end{equation}
Eqs.\ (\ref{fragpi}) and (\ref{fragmu})
give the decay products from charged pion and muon decays,
respectively. For simplicity, we adopt the notation $\pi$ for
$\pi^{\pm}$ and $e$ for $e^{\pm}$.

The source spectrum $q_{e}$ which is the number of electrons per unit time, volume, and
energy can be written as
\begin{equation}
q_e = n_{\chi}^2 \langle \sigma v \rangle
\left(\frac{dN_e}{dE_e}\right) \ ,
\label{qe}
\end{equation}
with $\langle \sigma v \rangle$ in this section denoting the thermally averaged cross section times 
velocity for annihilation into charged pions.
For electron energies relevant to the frequency range that we use, 
the source spectrum can be altered by synchrotron losses and inverse Compton scattering (ICS)
(see, e.g.,~\cite{blasi_colafrancesco99}).  The final spectrum 
$dn_{e}/ dE_{e}$ including losses is
\begin{equation}
\frac{dn_e}{dE_e} \simeq q_e \tau \,
\label{dndE}
\end{equation}
where $\tau$ is the average lifetime of an electron with energy $E_{e}$,
\begin{equation}
\tau \simeq {E_e \over dE_e/dt}\,.
\label{lifetime}
\end{equation}
The dominant loss process and, thus, the shortest lifetime, is used in
Eq.\ (\ref{dndE}). 

The total synchrotron power emitted from an electron of energy $E_{e}$, in the
presence of a magnetic field $B_{\mu} = | \vec{B} |$ in microgauss, 
is~\cite{rybicki_lightman}
\begin{equation}
{dE_{{\rm syn}} \over dt} = 1.6 \times 10^{-15} B_{\mu}^2 E_e^2
\,\,\,{\rm erg/s}\ .
\label{dEdtsyn}
\end{equation}
Combining Eqs.\ (\ref{lifetime}) and (\ref{dEdtsyn}) 
we find $\tau_{syn}$, the time scale for energy losses due
to synchrotron emission. 

In order to estimate $\tau_{syn}$ one needs the local magnetic field $B_{\mu}$. There are numerous estimates in the literature of
the component of the magnetic field along the line of sight $B_{\mu, los}$. Rotation measure 
analyses of pulsars in the LMC indicate a $B_{\mu, los}$ ranging from $\sim$ 0.4 to 5 $\mu G$~\cite{costa_etal91,costa_paper_etal91}.
Similar values ($B_{\mu,los} \simeq 2\ \mu G$) were obtained from the
polarized radio continuum emission~\cite{klein_etal93}.  Estimates of the 
total magnetic field, rather than the
component of the magnetic field along the line of sight, are more indirect and 
typically involve more assumptions.  Several studies seem to
converge to $\sim 6\ \mu G$ ~\cite{klein_etal89,haynes_etal91},
while the largest value estimated for the total magnetic field is $\simeq 18.4\ 
\mu G$~\cite{chi_wolfendale93}. 
In what follows we adopt the value 
of $\simeq 5\ \mu G$ for the total magnetic field, unless otherwise specified.
 
For magnetic fields of 5 $ \mu G$, synchrotron losses dominate over 
 ICS processes off the cosmic microwave background, the optical photons, and the produced synchrotron photons
 themselves. Therefore, the relevant timescale in Eq.\ (\ref{dndE}) is $\tau_{syn}$.

The synchrotron emission coefficient $j_{\nu}$ (number of photons of frequency $\nu$ emitted per unit time, volume, and frequency) is 
\begin{equation}
j_{\nu} = \frac{dn_e}{dE_e} \frac{dE_e}{d\nu} \frac{dE_{\rm syn}}{dt}
\,\,\, \frac{{\rm erg}}{{\rm cm^3\,s\,Hz}} \,.
\label{jnu}
\end{equation}
To calculate the factor $dE_{e}/d \nu$ appearing in
the expression for the emission coefficient,
we use the relation between $E_{e}$ and the frequency where maximum
synchrotron emission occurs for this electron energy. 
The  frequency of maximum synchrotron emission is 
equal to $\simeq 0.29 \nu_{c}$, 
with the  cutoff frequency $\nu_{c}= 3/2 \gamma^{3} \nu_{B}$ where 
$\gamma$ is the Lorentz
factor and $\nu_{B}=(eB \sin{\alpha} /2\pi \gamma m c)$ is the 
relativistic gyrofrequency.   
Taking the power weighted average over the pitch angle $\alpha$, the
frequency of maximum emission is
\begin{equation}
\nu_{{\rm max}} = 4.1\  B_{\mu} \left (\frac{E_e}{GeV} \right )^2 \ \rm{MHz} \ .
\label{numax}
\end{equation}
Using  Eqs.\ (\ref{fragtot}), (\ref{qe}), (\ref{dndE}), (\ref{lifetime}), (\ref{dEdtsyn}), and (\ref{numax}) to calculate 
$j_{\nu}$ from Eq.\ (\ref{jnu}), and substituting the latter in Eq.\ (\ref{fluxeqn}) 
we calculate the synchrotron flux. For the integration we use the same $R$ that
we use in the
$\gamma$-rays since the extent of the LMC in the two frequency
ranges is similar~\cite{sreekumar_etal92}.

The synchrotron flux 
$F_{syn}$ in Jy (1 Jy=$10^{-23}$ ergs cm$^{-2}$ s$^{-1}$ Hz$^{-1}$) is
\begin{equation}
F_{syn} \simeq 3.1 \times 10^{10} \frac{\langle \sigma v \rangle}{m^{2}_{\chi}}
\frac{dN_e}{dE_e} \frac{K}{B_{\mu}} \ \rm{Jy} \ ,
\end{equation}
where  $K$ is given by Eq.\ (\ref{K}) and is in units of 
(GeV/c$^2$)$^2$ cm$^{-5}$. Again, the neutralino mass  $m_{\chi}$ is given in GeV and $\langle \sigma v \rangle$ in cm$^3$ s$^{-1}$.   
The frequency dependence of the flux is given by 
\begin{eqnarray}
\frac{dN_e}{dE_e} \propto 2.48-1.22x^{-0.5} +0.22x^{2}+ 0.10x^{-1.5} 
 -1.54x^{0.5} -0.04x^{3} \ , 
\end{eqnarray}
where $x=E_e/m_{\chi} c^{2} \simeq (4.94 \times {10^{-4}} {B_{\mu}^{-0.5}} \nu^{0.5} / m_{\chi})$, with
$\nu$ in Hz and $m_{\chi}$ in GeV. 
The dependence on $\nu$ varies with the electron energy and the neutralino mass and  goes as $\nu^{-0.75}$ at sufficiently small energies. 

\begin{figure}[htb]
\vspace{2cm}
\centerline{\psfig{figure=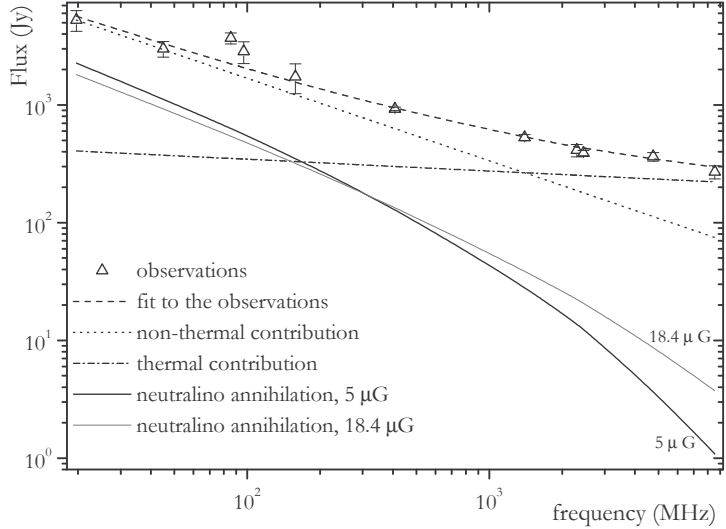,width=4.5in}}
\setlength{\unitlength}{0.0105in}%
\vspace{-3cm}
\caption{The flux in Jy as a function of frequency in the range from 19.7 to 8550 MHz. The open triangles
correspond to LMC data taken over the period from 1959 to 1991; the
dashed line is a best fit to the data assuming a decomposition into thermal and non-thermal emission. 
The dotted line corresponds to the non-thermal contribution, and the dot-dashed line to the thermal contribution. 
The flux from neutralino annihilation for $m_{\chi}=$ 50 GeV, $B_{\mu}= 5\ \mu G$, and
$\langle \sigma v \rangle \simeq 2 \times 10^{-26}$ cm$^3$ s$^{-1}$, is given by the thick solid line. 
The remaining line corresponds to $B_{\mu}= 18.4\ \mu G$, which is the maximum published estimate for
the total magnetic field (see text for details).  The $B_{\mu} \simeq 5\ \mu G$ case is the most realistic
value, even though the  results in these frequencies do not change dramatically  even when a
$18.4\ \mu G$ field is used. }
\label{results}

\end{figure}

A compilation of radio observations of the LMC from 1959 to 1991 in the frequency range $(19.7- 8550)$ MHz is presented along with the corresponding references 
in~\cite{haynes_etal91}. As can be seen in Fig.\ \ref{results},
the observed LMC spectrum in the  $(19.7-8550)$ MHz  range 
has a steep contribution from synchrotron radiation which dominates at lower frequencies until around $2 \times 10^{3}$ MHz,
at which point the thermal emission becomes dominant. A least squares fit of the form $S_{\nu} = S_{0,th} \nu^{-0.1} + S_{0,nth} \nu^{-\alpha_{nth}}$ to the data yields a total flux density at 1 GHz  of $610 \pm 10$ Jy, a non-thermal spectrum with $\alpha_{nth} =0.70 \pm 0.06$, and a fraction of thermal emission at 1 GHz of $(45 \pm 10) \%$~\cite{haynes_etal91}.  Using this information we find $S_{0,th} \simeq 2.2 \times 10^3$ and $S_{0,nth} \simeq 6.7 \times 10^{8}$. 
The data points and the decomposition into thermal and non-thermal emission are shown in Fig. \ref{results}. We also plot $F_{syn}$ for $m_{\chi}=50$ GeV and  $\langle \sigma v \rangle= 2.2 \times 10^{-26}$  cm$^3$ s$^{-1}$, for 
the NFW dark matter profile. The thick solid line corresponds to $ B_{\mu}=5 \mu G$; 
the lighter line represents the results for the maximum total magnetic
field  estimated in literature (18.4 $\mu G$~\cite{chi_wolfendale93}). 

In the more realistic case of $B_{\mu}=5\ \mu G$, neutralino-induced synchrotron emission clearly may be part of the observed flux, 
but is lower than the total observed flux, especially at higher frequencies.
In the low frequency region, the signal increases enough to exceed the
thermal contribution to the observation. 
Given that the neutralino signal ($\propto \nu^{-0.75}$) is almost 
parallel to the non-thermal component  ($\propto \nu^{-0.70}$), even 
at reasonably low radio frequencies the neutralino flux is sub-dominant. A mildly constraining limit on neutralino cross-sections can be reached by requiring that $F_{syn}$ 
be less than or equal to the non-thermal contribution in $S_{\nu}$ at low frequencies.  The frequency  dependence essentially drops out and we find  $\langle \sigma v \rangle \leq 
1.24 \times 10^{-9} m_{\chi}^{1.5} B_{\mu}^{0.25}/K$. 
Thus, assuming the NFW fit and $m_{\chi} =50$ GeV,   
any $\langle \sigma v \rangle$ lower than $\sim 2 \times 10^{-26}$ cm$^3$ s$^{-1}$ gives a flux at low frequencies consistent
with observation. The allowed $\langle \sigma v \rangle$
become even larger at larger neutralino masses. Thus,  the  observations 
do not put severe constraints on the allowed cross sections.
Although we have focused on the NFW profile, the synchrotron fluxes vary slightly depending on the halo profile used, as discussed in 
\S III.

The best hope for distinguishing the synchrotron emission generated by cosmic rays from that produced by neutralino annihilation
is to use the fact that the density profile of cosmic rays differs from the dark matter density profile.  While the dark matter halo extends significantly beyond the disk, the density of cosmic rays is expected to rapidly decrease at large radii.  If low radio frequency observations of the LMC are made with high angular resolution, the change from cosmic ray to neutralino dominance should be apparent as one moves away from the LMC disk. Currently there is no telescope that
could carry out observations of the LMC at frequencies less than about $20$ MHz.
Such low frequencies are extremely difficult to observe from the ground due to ionospheric
absorption and scattering. One promising ground-based 
project is the Low Frequency Array (LOFAR)~\cite{rottgering03} which should reach 
frequencies  down to $\sim$ 10 MHz and a flux sensitivity of a few $m$Jy  in 1 hour. 
However, the site of the instrument has not yet been 
decided, so the LMC may not necessarily be observable. 
To reach even lower frequencies, where ionospheric  absorption is very intense,
space-based instruments are required. The proposed Astronomical Low Frequency Array (ALFA)~\cite{jones_etal00} should reach down to $\sim 0.3$ MHz.

\section{Discussion and conclusions}
We have studied the $\gamma$-ray 
and synchrotron emission from neutralino annihilation in the LMC. We 
modeled the dark matter profile of the LMC by fitting 
 modern profiles found in N-body simulations to rotation velocity data. Although 
we focused on the NFW best fit density profile when deriving 
our $\gamma$-ray and synchrotron fluxes, we also considered a 
less concentrated dark matter profile (the isothermal sphere with a 
core) 
and a more concentrated one (the Moore et al.\  profile). 
This range of profiles 
changes our results by less than an order of magnitude for the same rotation curve.
If one takes into account the expected compression of the dark matter due to baryons, which
would cause even a shallow, core-dominated dark matter distribution to become cuspy, it is reasonable to assume
that the actual fluxes may be several orders of magnitude greater.
Prada et al.\ \cite{prada_etal04} found that baryonic compression can
enhance the central density  by more than an order of magnitude.

The LMC rotation curve data has been recently 
re-analyzed by  van der Marel et al.\ \cite{vandermarel_etal02} who claim that the plateau value for the rotation velocity is not 
$\sim$ (60-70) km/s as found in most studies, 
but $\sim$ 50 km/s. Their conclusion is based 
on a re-analysis of carbon star data including corrections 
for the transverse motion of the LMC, as well 
as for nutation and precession effects. Their 
resulting rotation curve has large errors and is sparse at small 
radii, but a fit to their rotation curve results in $K$ up 
to  $\sim$ 1 order of magnitude less than those we 
found using the initial data.  The quality of the fits is much poorer given the sparseness at small radii.

An additional uncertainty is introduced by the fact that only the line-of-sight component of
the velocity can be observed. The assumptions necessary to recover the total velocity from the
line-of-sight component may result in an over-- or underestimation of the actual velocity, thereby 
improving or worsening the prospects for detection. 

The predicted cosmic ray induced $\gamma$-ray and synchrotron emission appears to be of similar magnitude
to the observed emission. A high 
angular resolution study of both neutralino and cosmic ray 
induced emission from the 
LMC may be key to utilizing the closest known dark matter ``clump'' as a laboratory for testing neutralino parameters.
Smaller solid angles help reduce the collection of noise, and,  
most importantly, a study of the spatial dependence of the emission 
can help GLAST and 
radio observations distinguish between neutralino and cosmic ray produced signals.

Summarizing, we find that neither the EGRET 
measurement of $\gamma$-rays above 100 MeV from the LMC nor the low frequency synchrotron observations significantly constrain the SUSY parameter space.
The maximum  $\gamma$-ray flux for the considered profiles is almost $\sim 2 $ orders of magnitude
less than the observed flux, whereas all models with $\langle \sigma v 
\rangle$ $\la$ $2 \times 10^{-26}$ cm$^3$ s$^{-1}$ are consistent with the observed synchrotron flux. 
Existing and upcoming ACTs will not be able to probe  the SUSY 
parameter space, unless the neutralino profiles are more centrally concentrated or 
observers use  longer integration times and larger effective areas larger than currently planned.  
Since the sensitivity becomes better  for a larger exposure (effective area 
$\times$ integration time), the problem 
of a small effective area may be compensated for 
by larger integration  times. However, these integration 
times may not be  realistic. Added to these difficulties, the location 
of the LMC makes it difficult to observe even for some instruments 
in the southern hemisphere.

GLAST on the other hand, being a satellite, will not find the LMC location 
problematic. If systematic errors are small, GLAST  
will  be able to probe a significant part  of the allowed 
parameter space, especially if  the recent limit $m_{\chi}< 500 $ GeV is taken into account. The comparison of the 
$\gamma$-ray annihilation signal to the EGRET measurement implies that there is another 
dominant source of this flux, most likely cosmic rays.
The synchrotron study finds this result as well.  Thus, the spectral features
of the neutralino annihilation signal, in particular the shape and cutoff
of the spectrum at $m_{\chi}$, will be essential to disentangle the neutralino signal.    

The synchrotron signal is at best a factor of $\sim 3$ 
less than the observed signal, with this discrepancy increasing 
with increasing frequency. 
The flux values for the frequency range we use vary from $\sim 1 $ to $\sim 1000$ Jy, 
easily detectable levels. The difficulty will again be in disentangling the neutralino induced signal from the total flux. 
At low frequencies, the frequency dependence of the calculated flux 
is almost identical to that of the observed spectrum.
The best hope for observing the synchrotron emission due to neutralino annihilation 
is a high angular resolution study at low frequencies which may be possible with proposed observatories such as LOFAR or ALFA.

\section*{Acknowledgements}
This work was supported in part by the 
NSF through grant AST-0071235 and DOEgrant DE-FG0291-ER40606 
at the University of Chicago and at the Center for Cosmological 
Physics by grant NSF PHY-0114422.  A.O.\ thanks the Aspen Center for Physics for
its hospitality while this work was completed. 

\bibliography{tasitsiomi_etal}

\begin{thebibliography}{10}
\expandafter\ifx\csname url\endcsname\relax
  \def\url#1{\texttt{#1}}\fi
\expandafter\ifx\csname urlprefix\endcsname\relax\def\urlprefix{URL }\fi

\bibitem{spergel_etal03}
D.~N. {Spergel}, L.~{Verde}, H.~V. {Peiris}, E.~{Komatsu}, M.~R. {Nolta}, C.~L.
  {Bennett}, M.~{Halpern}, G.~{Hinshaw}, N.~{Jarosik}, A.~{Kogut}, M.~{Limon},
  S.~S. {Meyer}, L.~{Page}, G.~S. {Tucker}, J.~L. {Weiland}, E.~{Wollack},
  E.~L. {Wright}, astro-ph/0302209.

\bibitem{jungman_etal96}
G.~{Jungman}, M.~{Kamionkowski}, K.~{Griest}, Phys. Rep. 267 (1996) 195.

\bibitem{ellis_etal84}
J.~{Ellis}, J.~S. {Hagelin}, D.~V. {Nanopoulos}, K.~{Olive}, M.~{Srednicki},
  Nuclear Physics B 238 (1984) 453.

\bibitem{ellis_etal03}
J.~{Ellis}, K.~A. {Olive}, Y.~{Santoso}, V.~C. {Spanos}, hep-ph/0303043.

\bibitem{hagiwara_02}
K.~{Hagiwara}, et~al., Phys. Rev. D 66 (2002) 10001.

\bibitem{berezinsky_etal92}
V.~{Berezinsky}, A.~{Gurevich}, K.~{Zybin}, Phys. Rev. Lett. B 294 (1992) 221.

\bibitem{gondolo_silk99}
P.~{Gondolo}, J.~{Silk}, Phys. Rev. Lett. 83 (1999) 1719.

\bibitem{cesarini_etal03}
A.~{Cesarini}, F.~{Fucito}, A.~{Lionetto}, A.~{Morselli}, P.~{Ullio},
  astro-ph/0305075.

\bibitem{baltz_etal00}
E.~A. {Baltz}, C.~{Briot}, P.~{Salati}, R.~{Taillet}, J.~{Silk}, Phys. Rev. D
  61 (2000) 23514.

\bibitem{tyler_02}
C.~{Tyler}, Phys. Rev. D 66 (2002) 23509.

\bibitem{vassiliev_03}
V.~V. {Vassiliev}, astro-ph/0305584.

\bibitem{bergstrom_etal99}
L.~{Bergstr{\" o}m}, J.~{Edsj{\" o}}, P.~{Gondolo}, P.~{Ullio}, Phys. Rev. D 59
  (1999) 43506.

\bibitem{calcaneo-roldan_moore00}
C.~{Calc{\' a}neo-Rold{\' a}n}, B.~{Moore}, Phys. Rev. D 62 (2000) 123005.

\bibitem{tasitsiomi_olinto02}
A.~{Tasitsiomi}, A.~V. {Olinto}, Phys. Rev. D 66 (2002) 83006.

\bibitem{ullio_etal02}
P.~{Ullio}, L.~{Bergstr{\" o}m}, J.~{Edsj{\" o}}, C.~{Lacey}, Phys. Rev. D 66
  (2002) 123502.

\bibitem{berezinsky_etal03}
V.~{Berezinsky}, V.~{Dokuchaev}, Y.~{Eroshenko}, astro-ph/0301551.

\bibitem{blasi_etal03}
P.~{Blasi}, A.~V. {Olinto}, C.~{Tyler}, Astropart. Phys. 18 (2003) 649.

\bibitem{taylor_silk03}
J.~E. {Taylor}, J.~{Silk}, Mon. Not. R. Astron. Soc. 339 (2003) 505.

\bibitem{gondolo_94}
P.~Gondolo, Nucl. Phys. Proc. Suppl. 35 (1994) 148.

\bibitem{falvard_etal02}
A.~{Falvard}, E.~{Giraud}, A.~{Jacholkowska}, J.~{Lavalle}, E.~{Nuss},
  F.~{Piron}, M.~{Sapinski}, P.~{Salati}, R.~{Taillet}, K.~{Jedamzik},
  G.~{Moultaka}, astro-ph/0210184.

\bibitem{giraud_etal02}
E.~{Giraud}, G.~{Meylan}, M.~{Sapinski}, A.~{Falvard}, A.~{Jacholkowska},
  K.~{Jedamzik}, J.~{Lavalle}, E.~{Nuss}, G.~{Moultaka}, F.~{Piron},
  P.~{Salati}, R.~{Taillet}, astro-ph/0209230.

\bibitem{colafrancesco_mele01}
S.~{Colafrancesco}, B.~{Mele}, Astrophys. J. 562 (2001) 24.

\bibitem{vandermarel_etal02}
R.~P. {van der Marel}, D.~R. {Alves}, E.~{Hardy}, N.~B. {Suntzeff}, Astron. J.
  124 (2002) 2639.

\bibitem{meatheringham_etal88}
S.~J. {Meatheringham}, M.~A. {Dopita}, H.~C. {Ford}, B.~L. {Webster},
  Astrophys. J. 327 (1988) 651.

\bibitem{schommer_etal92}
R.~A. {Schommer}, N.~B. {Suntzeff}, E.~W. {Olszewski}, H.~C. {Harris}, Astron.
  J. 103 (1992) 447.

\bibitem{kim_etal98}
S.~{Kim}, L.~{Staveley-Smith}, M.~A. {Dopita}, K.~C. {Freeman}, R.~J. {Sault},
  M.~J. {Kesteven}, D.~{McConnell}, Astrophys. J. 503 (1998) 674.

\bibitem{sofue_99}
Y.~{Sofue}, Pub. Astron. Soc. Jap. 51 (1999) 445.

\bibitem{alves_nelson00}
D.~R. {Alves}, C.~A. {Nelson}, Astrophys. J. 542 (2000) 789.

\bibitem{navarro_etal95}
J.~F. {Navarro}, C.~S. {Frenk}, S.~D.~M. {White}, Mon. Not. R. Astron. Soc. 275
  (1995) 720.

\bibitem{navarro_etal96}
J.~F. {Navarro}, C.~S. {Frenk}, S.~D. {White}, Astrophys. J. 462 (1996) 563.

\bibitem{moore_etal98}
B.~Moore, F.~Governato, T.~Quinn, J.~Stadel, G.~Lake, Astrophys. J. 499 (1998)
  L5.

\bibitem{tasitsiomi_02}
A.~Tasitsiomi, Int. J. Mod. Phys. D 12 (2003) 1157.

\bibitem{prada_etal04}
F.~{Prada}, A.~{Klypin}, J.~{Flix}, M.~{Martinez}, E.~{Simonneau},
  {Astrophysical inputs on the SUSY dark matter annihilation detectability},
  ArXiv Astrophysics e-prints.

\bibitem{hayashi_etal03}
E.~{Hayashi}, J.~F. {Navarro}, J.~E. {Taylor}, J.~{Stadel}, T.~{Quinn},
  Astrophys. J. 584 (2003) 541.

\bibitem{kravtsov_03}
A.~V. Kravtsov, private communication.

\bibitem{klypin_etal99}
A.~{Klypin}, S.~{Gottl{\" o}ber}, A.~V. {Kravtsov}, A.~M. {Khokhlov},
  Astrophys. J. 516 (1999) 530.

\bibitem{stoehr_etal02}
F.~{Stoehr}, S.~D.~M. {White}, G.~{Tormen}, V.~{Springel}, Mon. Not. R. Astron.
  Soc. 335 (2002) L84.

\bibitem{hill_83}
C.~T. {Hill}, Nucl. Phys. B 224 (1983) 469.

\bibitem{bergstrom_etal01}
L.~{Bergstr{\" o}m}, J.~{Edsj{\" o}}, P.~{Ullio}, Phys. Rev. Lett. 87 (2001)
  251301.

\bibitem{sreekumar_etal92}
P.~{Sreekumar}, D.~L. {Bertsch}, B.~L. {Dingus}, C.~E. {Fichtel}, R.~C.
  {Hartman}, S.~D. {Hunter}, G.~{Kanbach}, D.~A. {Kniffen}, Y.~C. {Lin}, J.~R.
  {Mattox}, H.~A. {Mayer-Hasselwander}, P.~F. {Michelson}, C.~{von Montigny},
  P.~L. {Nolan}, K.~{Pinkau}, E.~J. {Schneid}, D.~J. {Thompson}, Astrophys. J.
  Lett. 400 (1992) L67.

\bibitem{hartman_etal99}
R.~C. {Hartman}, et~al., Astrophys. J. Suppl. Ser. 123 (1999) 79.

\bibitem{pavlidou_fields01}
V.~{Pavlidou}, B.~D. {Fields}, {Diffuse Gamma Rays from Local Group Galaxies},
  Astrophys. J. 558 (2001) 63--71.

\bibitem{sreekumar_etal98}
P.~{Sreekumar}, D.~L. {Bertsch}, B.~L. {Dingus}, J.~A. {Esposito}, C.~E.
  {Fichtel}, R.~C. {Hartman}, S.~D. {Hunter}, G.~{Kanbach}, D.~A. {Kniffen},
  Y.~C. {Lin}, H.~A. {Mayer-Hasselwander}, P.~F. {Michelson}, C.~{von
  Montigny}, A.~{Muecke}, R.~{Mukherjee}, P.~L. {Nolan}, M.~{Pohl},
  O.~{Reimer}, E.~{Schneid}, J.~G. {Stacy}, F.~W. {Stecker}, D.~J. {Thompson},
  T.~D. {Willis}, Astrophys. J. 494 (1998) 523.

\bibitem{bergstrom_etal98}
L.~{Bergstr{\" o}m}, P.~{Ullio}, J.~H. {Buckley}, Astropart. Phys. 9 (1998)
  137.

\bibitem{longair_92}
M.~Longair, High Energy Astrophysics, Cambridge University Press, Cambridge,
  England, 1992.

\bibitem{DarkSUSY}
P.~{Gondolo}, J.~{Edsjo}, P.~{Ullio}, L.~{Bergstrom}, M.~{Schelke}, E.~A.
  {Baltz}, astro-ph/0211238.

\bibitem{deangelis_00}
A.~de~Angelis, astro-ph/0009271.

\bibitem{hofmann_etal00}
W.~{Hofmann}, {The Hess Collaboration}, {The High Energy Stereoscopic System
  (HESS) Project}, in: GeV-TeV Gamma Ray Astrophysics Workshop : towards a
  major atmospheric Cherenkov detector, 2000, p. 500.

\bibitem{mori_etal00}
M.~{Mori}, et~al., {The CANGAROO-III Project}, in: GeV-TeV Gamma Ray
  Astrophysics Workshop : towards a major atmospheric Cherenkov detector, 2000,
  p. 485.

\bibitem{blasi_colafrancesco99}
P.~{Blasi}, S.~{Colafrancesco}, Astropart. Phys. 12 (1999) 169.

\bibitem{rybicki_lightman}
G.~{Rybicki}, A.~{Lightman}, Radiative Processes in Astrophysics, Wiley, New
  York, 1979.

\bibitem{costa_etal91}
M.~E. {Costa}, P.~M. {McCulloch}, P.~A. {Hamilton}, {The Magnetic Field
  Strength in the Large Magellanic Cloud}, in: IAU Symp. 148: The Magellanic
  Clouds, 1991, p. 101.

\bibitem{costa_paper_etal91}
M.~E. {Costa}, P.~M. {McCulloch}, P.~A. {Hamilton}, Mon. Not. R. Astron. Soc.
  252 (1991) 13.

\bibitem{klein_etal93}
U.~{Klein}, R.~F. {Haynes}, R.~{Wielebinski}, D.~{Meinert}, Astron. Astrophys.
  271 (1993) 402.

\bibitem{klein_etal89}
U.~{Klein}, R.~{Wielebinski}, R.~F. {Haynes}, D.~F. {Malin}, Astron. Astrophys.
  211 (1989) 280.

\bibitem{haynes_etal91}
R.~F. {Haynes}, U.~{Klein}, S.~R. {Wayte}, R.~{Wielebinski}, J.~D. {Murray},
  E.~{Bajaja}, D.~{Meinert}, U.~R. {Buczilowski}, J.~I. {Harnett}, A.~J.
  {Hunt}, R.~{Wark}, L.~{Sciacca}, Astron. Astrophys. 252 (1991) 475.

\bibitem{chi_wolfendale93}
X.~{Chi}, A.~W. {Wolfendale}, Nature 362 (1993) 610.

\bibitem{rottgering03}
H.~{Rottgering}, A.~G. {de Bruyn}, R.~P. {Fender}, J.~{Kuijpers}, M.~P. {van
  Haarlem}, M.~{Johnston-Hollitt}, G.~K. {Miley}, astro-ph/0307240.

\bibitem{jones_etal00}
D.~L. {Jones}, et~al., Advances in Space Research 26 (2000) 743.

\end{thebibliography}

\end{document}